\documentclass[10pt, conference]{IEEEtran}
\usepackage{graphicx}
\usepackage{mhchem}
\usepackage{subfig}
\usepackage{url}
\usepackage{epstopdf}
\usepackage{amsmath}
\usepackage{amssymb}
\usepackage{balance}
\usepackage{cite}
\usepackage{citesort}
\usepackage[usenames, dvipsnames]{color}

\begin{document}
\bstctlcite{IEEEexample:BSTcontrol}

%0.67 (L) x 0.67 (R) x 0.81 (T) x 0.92 (B)

\title{A New Look at MIMO Capacity \\
in the Millimeter Wave}

\author{\IEEEauthorblockN{Sayed Amir Hoseini\IEEEauthorrefmark{1}\IEEEauthorrefmark{2}, Ming Ding\IEEEauthorrefmark{2} and Mahbub Hassan\IEEEauthorrefmark{1}\IEEEauthorrefmark{2}} 	\\
	{\IEEEauthorrefmark{1}School of Computer Science and Engineering, University of New South Wales, Sydney,  Australia} \\
	{\IEEEauthorrefmark{2}Data61, CSIRO, Sydney, Australia}\\
		Email: s.a.hoseini@unsw.edu.au, Ming.Ding@data61.csiro.au, mahbub.hassan@unsw.edu.au}

\maketitle

\begin{abstract}
In this paper,
we present a new theoretical discovery that the multiple-input and multiple-output (MIMO) capacity can be influenced by atmosphere molecules.
In more detail,
some common atmosphere molecules,
such as Oxygen and water,
can absorb and re-radiate energy in their natural resonance frequencies,
such as 60\,GHz, 120\,GHz and 180\,GHz,
which belong to the millimeter wave (mmWave) spectrum.
Such phenomenon can provide equivalent non-line-of-sight (NLoS) paths in an environment that lacks scatterers,
and thus greatly improve the spatial multiplexing and diversity of a MIMO system.
This kind of performance improvement is particularly useful for most mmWave communications that heavily rely on line-of-sight (LoS) transmissions.
%Our results show that some frequency bands the MIMO capacity of LoS channel can reach to the theoretical maximum.
To sum up,
our study concludes that since the molecular re-radiation happens at certain mmWave frequency bands,
%such as 60GHz, 120GHz and 180GHz,
the MIMO capacity becomes highly \emph{frequency selective} and enjoys a considerable boosting at those mmWave frequency bands.
%How to exploit those frequency windows by means of MIMO,
%becomes an intriguing research direction for both scientists and engineers.
The impact of our new discovery is significant,
which fundamentally changes our understanding on the relationship between the MIMO capacity and the frequency spectrum.
In particular,
our results predict that several mmWave bands can serve as valuable spectrum windows for high-efficiency MIMO communications,
which in turn may shift the paradigm of research, standardization, and implementation in the field of mmWave communications.

\end{abstract}

\section{ Introduction}\label{sec:introduction}

In the near future,
the 5th-generation (5G) system is expected to deliver a massive increase in channel capacity and data rates.
To achieve this,
two key technologies have attracted lots of attention recently.
The first one is the use of very high frequency spectrum in the range of 30\,GHz to 300\,GHz,
which is also known as the millimeter wave (mmWave) spectrum.
The second one is the massive multiple-input multiple-output (MIMO) technology,
which advocates the use of a large number of antennas in wireless communications.
The above two technologies are compatible with each other.
%thanks to the small-size antennas in mmWave.
In more detail,
a short wavelength in mmWave helps to minimize the inter-element spacing of a MIMO system.
As a result,
a great number of antennas can be equipped on mobile devices,
which is not practical for the current wireless systems that usually work in the sub-6GHz spectrum.

For a given number of transmitter and receiver antennas,
the common understanding is that the MIMO capacity is not frequency selective,
if the received signal strength is fixed to a certain level.
%In spite of the possibility of placing more antennas in higher frequency,
%the MIMO capacity is not frequency selective for a fixed number of antennas.
%It means for invariant received signal to noise ratio the ideal MIMO channel capacity will not change with the frequency.
Such conclusion has been validated in the sub-6GHz spectrum,
\emph{but does this conclusion still hold when we march into the mmWave spectrum in 5G?}

Note that the mmWave communication for cellular networks poses its own challenges,
such as high free-space path loss, high Doppler shift and blockage~\cite{Rangan2014mmWaveSurvey}.
%which are currently under investigation.
Other than the above factors and according to more fundamental physics theories,
another key difference between the existing wireless communication in sub-6GHz frequencies and the future one in mmWave is the reaction of atmosphere molecules,
which can absorb
%and re-radiate
signal energy if excited in their natural resonance frequencies.
Such natural resonance frequencies are usually in the mmWave spectrum.
For example,
if we consider normal atmosphere,
Oxygen and/or water molecules will play a major role in the molecular absorption,
and their natural resonance frequencies are around 60\,GHz, 120\,GHz and 180\,GHz.
Interestingly,
resonating Oxygen and water molecules not only absorb signal energy causing attenuation,
they also re-radiate some of the absorbed energy.
%and thus sending distorted and delayed signals to the receiver.
This type of molecule-induced re-radiation is often referred to as \emph{molecular noise}~\cite{Akyildiz201416,Noise_1986},
but it is actually highly correlated to the signal waveform due to its re-radiation nature~\cite{jornet2013fundamentals},
and hence it can be considered as a distorted copy of the signal from a virtual non-line-of-sight (NLoS) path.

The aim of this paper is to show how this molecular re-radiation can change the MIMO capacity performance as it is very similar to scattering,
which is known as a very important factor to provide spatial diversity for a MIMO channel.
Since the molecule absorption intensity is related to the natural resonance frequencies of the existing molecules in the realistic atmosphere on earth,
(i.e., mainly the large amount of Oxygen and water molecules),
the energy absorption and re-radiation are not flat especially in the mmWave spectrum.
Thus,
we propose a conjecture that the MIMO capacity should vary with frequency due to the impact of such molecular absorption and re-radiation.
Then,
we verify our conjecture via theoretical studies and computer simulations,
and find that the MIMO capacity increases dramatically in some high absorption bands around 60\,GHz, 120\,GHz and 180\,GHz,
thanks to the ubiquitous existence of Oxygen and water molecules.

The intuition of our theoretical discovery is that the molecular re-radiation adds a random phase onto the distorted copy of the signal,
which equivalently creates a richer scattering environment that can improve the line-of-sight (LoS) MIMO capacity commonly evaluated in the mmWave spectrum.
%provide a richer scattering environment which can improve the line of sight communication capacity in free space.
%We find the molecules re-radiation with a random phase provide a richer scattering environment which can improve the line of sight communication capacity in free space.
The impact of our new discovery is significant,
which fundamentally changes our understanding on the relationship between the MIMO capacity and the frequency spectrum.
In particular,
our results predict that several mmWave bands can serve as valuable spectrum windows for high-efficiency MIMO communications,
which in turn may shift the paradigm of research, standardization, and implementation in the field of mmWave communications.

The rest of the paper is structured as follows.
%In Section~\ref{sec:Related}, 
%we briefly review the related work.
In Section~\ref{sec:model}, 
we present the molecular absorption model for the calculation of attenuation and re-radiation,
as well as the fundamental theory on the MIMO capacity.
Section~\ref{sec:analysis} analyzes the MIMO capacity versus the molecular re-radiation, 
followed by simulation results and insightful discussion in Section~\ref{sec:simulSetup}.
Finally,
we conclude the paper in Section~\ref{sec:con}.

%%%%%%%%%%%%%%%%%%%%%%%%%%%%%%%%%%%%%%%%%%%%%%%%%%%%%%%%%%%%%%%%%%%%%%%%%
%%%%%%%%%%%%%%%%%%%%%%%%%%%%%%%%%%%%%%%%%%%%%%%%%%%%%%%%%%%%%%%%%%%%%%%%%
%
%\section{ Related Work}\label{sec:Related}
%
%The MIMO capacity has been well studied in the literature from different aspects,
%such as spatial diversity and correlation~\cite{chinani2003MIMOcap},
%array geometry optimization~\cite{sarris2005maximum},
%mutual coupling~\cite{kildal2004correlation},
%and so on.
%On the other hand,
%the molecular absorption and re-radiation are well-known phenomena,
%and they have been recently taken into account in mmWave and Terahertz research~\cite{Noise_1986,jornet2013fundamentals,Akdeniz2014Capacity,Akyildiz2010c}.
%While most of these works considered the re-radiated energy as noise,
%the authors in~\cite{jornet2013fundamentals} proposed that the self-induced noise should be correlated to the original signal,
%and it was modelled as multiple scattering from virtual NLoS paths~\cite{Kokkoniemi2015discuss}.
%However,
%to the best of our knowledge,
%there is no previous work on studying the impact of the molecular absorption and re-radiation on the MIMO capacity.
%%%%%%%%%%%%%%%%%%%%%%%%%%%%%%%%%%%%%%%%%%%%%%%%%%%%%%%%%%%%%%%%%%%%%%%%%
%%%%%%%%%%%%%%%%%%%%%%%%%%%%%%%%%%%%%%%%%%%%%%%%%%%%%%%%%%%%%%%%%%%%%%%%%

\section{Channel model and MIMO capacity} \label{sec:model}

The molecular absorption model defines how different species of molecules in a communication channel absorb energy from the electromagnetic signals and how they re-radiate them back to the environment.
This section first explains the concept of \emph{absorption coefficient} used to characterize the absorption capacity of a given molecule species,
followed by the attenuation and re-radiation models that are built upon this coefficient.

\subsection{Molecular absorption coefficient}

The effect of a given molecule,
denoted by $S_i$,
on the radio signal is characterized by its molecular absorption coefficient $K_i(f)$ at frequency $f$.
Such coefficient varies with pressure and temperature of the environment.
The molecular absorption coefficients of many chemical species for different pressure and temperature are available from the publicly available databases such as \emph{HITRAN}~\cite{Rothman2012Database} and \textit{NIST Atomic Spectra}~\cite{NIST}.
Nevertheless,
the atmospheric air is a mixture of different molecule species that may change on an hourly basis~\cite{hoseini2016diurnal}.
Therefore,
different climate conditions also lead to different absorption in mmWave and they should be taken into account when considering mmWave channels.
In order to model molecular absorption,
let us assume the mmWave radio channel is a medium consisting of $N$ chemical species $S_1,S_2, ..., S_N$,
and $m_i$ is the mole fraction per volume,
i.e., the mixing ratio of molecule $S_i$ in the channel medium.
We further assume that the temperature and pressure of the medium are $\mathcal{T} $ and $\mathcal{P}$, respectively.
The \emph{medium absorption coefficient},
i.e., $k(f)$,
at frequency $f$ is therefore a weighted sum of the molecular absorption coefficients in the medium~\cite{Jornet2014a}, which can be formulated as
\begin{equation}
k(f) = \sum_{i=1}^{N} m_i k_{i}(f),
\label{eq:Kf}
\end{equation}
where $k_{i}(f)$ is the molecular absorption coefficient of species $S_i$ on condition of temperature $\mathcal{T}$ and
and pressure $\mathcal{P}$.
As discussed before,
$k_{i}(f)$ can be obtained from HITRAN~\cite{Rothman2012Database} and NIST~\cite{NIST}.
In this work,
to get the values of $k(f)$,
we will use some predefined standard atmosphere conditions and their corresponding ratio of molecules in the air,
which are tabulated in~\cite{Rothman2012Database}.

\subsection{Attenuation of radio signal}

The attenuation of the radio signal at the mmWave frequencies is due to spreading and molecular absorption~\cite{mmWavepathloss}.
The total attenuation at frequency $f$ and a distance $d$ from the radio transmitter can be written as
\begin{equation}
A(f,d)  = A_{\rm spread}(f, d) \times A_{\rm abs}(f,d),
\label{eq:Atten_Total}
\end{equation}
where $A_{\rm spread}(f,d)$ and $A_{\rm abs}(f,d)$ are respectively the attenuation due to spreading and attenuation caused by molecular absorption at frequency $f$.

In more detail,
the spreading attenuation is given by
\begin{equation}
 A_{\rm spread}(f, d)  =  \left( \frac{4 \pi f d}{c}\right) ^2,
 \label{eq:Atten_Spread}
\end{equation}
where $c$ is the speed of light.
The attenuation due to molecular absorption is characterized as~\cite{mmWavepathloss}
\begin{equation}
A_{\rm abs}(f,d) =  e^{k(f) \times d}.
\label{eq:Atten_abs}
\end{equation}

Many linear approximations for~\eqref{eq:Atten_Total} have been proposed in the literature for different frequencies (in dB or dB/Km) \cite{mmWavePathloss1}.
The ITU Radio communication Sector (ITU-R) also provides various procedures to estimate specific attenuation due to the dominant molecules in the air (i.e., Oxygen and water)~\cite{P67610},
both of which can be derived from~\eqref{eq:Atten_Spread} and \eqref{eq:Atten_abs}.

Thus,
the line-of-sight (LoS) received power at the receiver becomes
\begin{eqnarray}
  \nonumber P_{{\rm r,LoS}}(f,d) &=& \frac{P_{t}(f)}{A(f,d)} \\
    &=& P_{t}(f)\times\left(\frac{c}{4\pi fd}\right)^{2}\times e^{-k(f)\times d}.
    \label{eq:Atten_pr1}
\end{eqnarray}

\subsection{Molecular re-radiation}

The existing molecules in communication medium will be excited by electromagnetic waves at specific frequencies.
The excitement is temporary and the vibrational-rotational energy level of molecules will come back to a steady state and the absorbed energy will be re-radiated in the same frequency.
These re-radiated waves are usually considered as noise in the literature.
For example,
the molecular absorption noise has been studied in the literature since 1986 when a model for sky atmospheric noise for frequencies higher than 18GHz~ was proposed in~\cite{Noise_1986}.
There are a number of works that have studied the atmospheric noise for mmWave frequencies such as~\cite{AtmosphericNoise_Mauna_Kea}, which experimentally measured the atmospheric noise variation in Mauna Kea in Hawaii over many night and days using a 143\,GHz and 268\,GHz transmitter.
Recently,
the molecular noise has been re-considered for higher frequencies such as Terahertz band ranging from 0.1-10\,THz~\cite{Akyildiz201416}.
Molecular absorption is not white and its power spectral density (PSD) is not flat because of the different resonant frequencies of various species of molecules.
The PSD of the molecular absorption noise that affects the transmission of a signal, $S_{N_{\rm abs}}$,
is contributed by the atmospheric noise $S^B_{N}$ and the self-induced noise $S^X_{N}$ as addressed in~\cite{Noise_1986,Jornet2014a}:
\begin{align}
&S_{N_{\rm abs}}(f,d)= S^B_{N}(f,d)+S^X_{N}(f,d), \label{eq:NoisePDS}\\
&S^B_{N}(f,d)=lim_{d\to\infty} (k_B T_0 (1-e^{-k(f) d})) \Big( \frac{c}{\sqrt{4\pi} f} \Big)^2, \label{eq:Noise01}\\
&S^X_{N}(f,d)=P_t(f)(1-e^{-k(f) d}) \Big(\frac{c}{{4\pi f d}}\Big)^2, \label{eq:Noise02}
\end{align}
where $k(f)$ is the absorption coefficient of the medium at frequency $f$,
$T_0$ is the reference temperature ($ 296K) $,
$k_B$ is the Boltzmann constant,
$P_t(f)$ is the power spectral density of the transmitted signal and $c$ is the speed of light.
The first term in~\eqref{eq:NoisePDS},
which is called sky noise and defined in~\eqref{eq:Noise01} is independent of the signal wave.
However,
the self-induced noise in~\eqref{eq:Noise02} is highly correlated with the signal wave~\cite{jornet2013fundamentals},
and can be considered as a distorted copy of the signal wave~\cite{Kokkoniemi2015discuss}.
Thus,
the received power of the re-radiated signal by molecules at the receiver can be expressed by
\begin{equation}
 P_{\rm r,a}(f,d)=P_t(f)(1-e^{-k(f) d}) \Big(\frac{c}{{4\pi f d}}\Big)^2.
\label{eq:Atten_pr2}
\end{equation}

Since the phase of the re-radiated wave depends on the phase of molecular vibration,
which varies from molecules to molecules~\cite{barron2004molecular},
the received power in this case is affected by a large number of phase-independent re-radiated photons.
Thus,
we assume a uniformly distributed \emph{random} phase $\beta_{\rm random}$, for the received signal, with its power given by~\eqref{eq:Atten_pr2}.

\subsection{Channel Transfer Function}

\label{sec:Hfunction}
The channel transfer function for a single LoS channel is given by
\begin{equation}
\begin{aligned}
H_{\rm LoS}(f,d) &=\sqrt{ \left( \frac{c}{4 \pi f d}\right) ^2 e^{-k(f) \times d}} \times e^ {j2\pi\frac{d}{\lambda}} \\
			   &= \left( \frac{c}{4 \pi f d}\right)  e^{-k(f) \times \frac{d}{2}} \times e^ {j2\pi\frac{d}{\lambda}}.
\label{eq:Hfunc_1}
\end{aligned}
\end{equation}

Then,
the partial channel transfer function resulted from the molecular absorption and excluding the LoS component can be represented by
\begin{equation}
\begin{aligned}
 H_{\rm a}(f,d)&=\sqrt{(1-e^{-k(f) d}) \Big(\frac{c}{{4\pi f d}}\Big)^2}\times e^ {j2\pi\beta_{\rm random}}\\
 			 &=(1-e^{-k(f) d})^{\frac{1}{2}} \Big(\frac{c}{{4\pi f d}}\Big) \times e^ {j2\pi\beta_{\rm random}}.
\label{eq:Hfunc_2}
\end{aligned}
\end{equation}
Hence,
the total channel transfer function is the superposition of the partial channel transfer functions,
which is written as
\begin{eqnarray}
  \nonumber H(f,d) \hspace{-0.2cm}&=&\hspace{-0.2cm} H_{\rm LoS}(f,d) + H_{\rm a}(f,d) \\
  \nonumber   \hspace{-0.2cm}&=&\hspace{-0.2cm} \left( \frac{c}{4 \pi f d}\right)  e^{-k(f) \times \frac{d}{2}} \times e^ {j2\pi\frac{d}{\lambda}} \\
    \hspace{-0.2cm}& &\hspace{-0.2cm} + (1-e^{-k(f) d})^{\frac{1}{2}} \Big(\frac{c}{{4\pi f d}}\Big) \times e^ {j2\pi\beta_{\rm random}}.
    \label{eq:Hfunc}
\end{eqnarray}

\subsection{MIMO channel model and capacity}

\begin{figure}[h]
\centering
\includegraphics[width=0.8\linewidth]{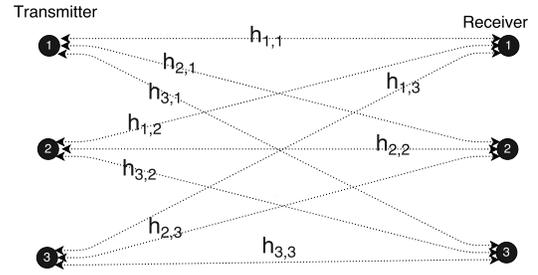}
\caption{A 3x3 MIMO system, the channel gain of array pairs between transmitter and receivers}
\label{fig:Hmatrix}
\end{figure}

In this paper,
we consider a MIMO system that is consisted of $n_t$ transmitting antennas and $n_r$ receiving ones.
The received signal vector $y$ at $n_r$ receiving antennas can be formulated as~\cite{chinani2003MIMOcap}
\begin{equation}
y=Hx+n,
\label{eq:MIMObase}
\end{equation}
where $x$ is the transmitted signal vector form $n_t$ transmitting antennas,
and $n$ is an $n_r\times1$ vector with zero-mean independent noises with variance $\sigma^2$.
The channel matrix $H$ is defined by
\begin{equation}
H \triangleq
\begin{bmatrix}
 &h_{11}  &h_{12}  & \ldots &h_{1n_t} \\
 &h_{21}  &h_{22}  & \ldots &h_{2n_t} \\
 &\vdots  &\vdots  & \ddots & \vdots  \\
 &h_{n_r1}  &h_{n_r2}  &\ldots &h_{n_rn_t},  \\
\end{bmatrix}
\label{eq:Hmatrix}
\end{equation}
where ${h_{ij}}$ is a complex value denoting the transfer coefficient associated with the $j$th transmitter antenna and the $i$th receiver antenna.
Note that ${h_{ij}}$ can be obtained from~\eqref{eq:Hfunc} for frequency $f$ and distance $d_{ij}$.
A $3\times3$ MIMO system with a channel matrix is illustrated in Figure~\ref{fig:Hmatrix}.

In this paper,
we assume that the transmitter has no channel state information (CSI),
and the transmitting power is equally distributed among transmitting antennas.
Consequently,
the capacity of MIMO channel can be written as
\begin{equation}
C = {\rm log}_{2}{\rm det} ({I}_{n_r}+\frac{P}{n_t\sigma^2}\mathbf{H}\mathbf{H}^\dagger),
\label{eq:MIMOcapBase}
\end{equation}
where $P$ is total transmitting power,
and $I$ is the identity matrix \cite{chinani2003MIMOcap}.
Since the determinant of $\left({I}_{n_r}+\frac{P}{n_t\sigma^2}\mathbf{H}\mathbf{H}^\dagger\right)$ can be computed by the product of the eigenvalues of the matrix $\mathbf{H}\mathbf{H}^\dagger$,
%and the zero eigenvalues of matrix $\mathbf{H}\mathbf{H}^\dagger$ do not contribute to such product,
the MIMO capacity can thus be written in the form of a product of non-zero eigenvalues as~\cite{Tse2005MIMObook}
\begin{equation}
C = \sum_{i=1}^{\kappa}{\rm log}_2(1+\frac{P\lambda_i^2}{n_t\sigma^2}),
\label{eq:MIMOcapEig}
\end{equation}
where $\lambda_i$ denotes singular values of the matrix $\mathbf{H}$,
and hence the squared singular values $\lambda_i^2$ denotes the eigenvalues of the matrix $\mathbf{H}\mathbf{H}^\dagger$.
Note that $\kappa$ denotes the number of non-zero $\lambda_i^2$,
which is also called the rank of $\mathbf{H}$ with $\kappa\leq {\rm min}(n_r,n_t)$~\cite{Tse2005MIMObook}.

%%%%%%%%%%%%%%%%%%%%%%%%%%%%%%%%%%%%%%%%%%%%%%%%%%%%%%%%%%%%%%%%%%%%%%%%%
%%%%%%%%%%%%%%%%%%%%%%%%%%%%%%%%%%%%%%%%%%%%%%%%%%%%%%%%%%%%%%%%%%%%%%%%%

\section{Analysis on the MIMO capacity with molecular absorption}
\label{sec:analysis}

\begin{figure*}[t]
% ensure that we have normalsize text
\normalsize
\begin{equation}
\begin{aligned}
\mathbf{H}\mathbf{H}^\dagger &= \left(\frac{c}{4 \eta \pi f}\right)^2
\\
&\times \begin{bmatrix}
 &\cfrac{e^{-k(f) \frac{d_{11}}{2}+ j2\pi\frac{d_{11}}{\lambda}}+ (1-e^{-k(f) d_{11}})^{1/_2}e^ {j2\pi\beta_{\rm n_{11}}}}{ d_{11}}   &\cfrac{e^{-k(f) \frac{d_{12}}{2}+ j2\pi\frac{d_{12}}{\lambda}}+ (1-e^{-k(f) d_{12}})^{1/_2}e^ {j2\pi\beta_{\rm n_{12}}}}{ d_{12}} \\
 %%%%%%%%%%%%%%%%%%%%%%%%%%%%%%%%%%%%%%%%%%%%%%%%%%%%
 &\cfrac{e^{-k(f) \frac{d_{21}}{2}+ j2\pi\frac{d_{21}}{\lambda}}+ (1-e^{-k(f) d_{21}})^{1/_2}e^ {j2\pi\beta_{\rm n_{21}}}}{ d_{21}}   &\cfrac{e^{-k(f) \frac{d_{22}}{2}+ j2\pi\frac{d_{22}}{\lambda}}+ (1-e^{-k(f) d_{22}})^{1/_2}e^ {j2\pi\beta_{\rm n_{22}}}}{ d_{22}} \\
\end{bmatrix} \\
&\times \begin{bmatrix}
 &\cfrac{e^{-k(f) \frac{d_{11}}{2}- j2\pi\frac{d_{11}}{\lambda}}+ (1-e^{-k(f) d_{11}})^{1/_2}e^ {-j2\pi\beta_{\rm n_{11}}}}{ d_{11}}   &\cfrac{e^{-k(f) \frac{d_{21}}{2}- j2\pi\frac{d_{21}}{\lambda}}+ (1-e^{-k(f) d_{21}})^{1/_2}e^ {-j2\pi\beta_{\rm n_{21}}}}{ d_{21}} \\
 %%%%%%%%%%%%%%%%%%%%%%%%%%%%%%%%%%%%%%%%%%%%%%%%%%%%
 &\cfrac{e^{-k(f) \frac{d_{12}}{2}- j2\pi\frac{d_{12}}{\lambda}}+ (1-e^{-k(f) d_{12}})^{1/_2}e^ {-j2\pi\beta_{\rm n_{12}}}}{ d_{12}}   &\cfrac{e^{-k(f) \frac{d_{22}}{2}- j2\pi\frac{d_{22}}{\lambda}}+ (1-e^{-k(f) d_{22}})^{1/_2}e^ {-j2\pi\beta_{\rm n_{22}}}}{ d_{22}} \\
\end{bmatrix}
\end{aligned}
\label{eq:HmatrixF}
\end{equation}
\end{figure*}

It is well known that the maximum achievable capacity of a MIMO channel is proportional to the minimum number of antenna elements at the receiver and the transmitter.
However,
in an environment that is lack of spatial diversity,
such capacity would be degraded due to the deficiency of parallel information paths,
i.e., the rank of the MIMO channel,
between the receiver and the transmitter~\cite{goldsmith2003capacity}.
More specifically,
in a rich scattering environment,
scatterers provide sufficient NLoS signal components,
leading to a better diversity and capacity.
But in the case of a LoS scenario,
the LoS signal component will dominate the received signal,
and thus decrease the channel rank due to the linear dependence of the LoS antenna array phases~\cite{sarris2005maximum}.
Thus,
in a LoS scenario that is assumed for most mmWave communications,
the maximum MIMO capacity is achievable only with some specific array configuration~\cite{sarris2005maximum},
where the LoS rays are perfectly orthogonal,
resulting in an opportunistically full-rank MIMO channel.
However,
this is not practical for mobile communications,
since such kind of optimal antenna setting requires a user steadily holding a device toward a specific direction.
%to have a fixed channel distance and very large inter-element spacing.

Generally,
the MIMO capacity becomes higher if the channel transfer matrix $\textbf{H}$ is \emph{full rank} and \emph{well-conditioned}.
In more detail,
the rank of $\textbf{H}$ determines how many data streams can be multiplexed over the channel,
and $\textbf{H}$ is well-conditioned if the condition number,
which is defined as $\frac{\lambda{max}}{\lambda{min}}$,
is small and close to one.
In other words,
the maximum MIMO capacity can be attained when all $\lambda_i$s are equal.

%To show that whether the absorption re-radiation can help to provide a scattering effect on LoS signal,
With the absorption re-radiation,
we can show that it helps to provide equivalent NLoS paths for the LoS scenario, 
and thus increase the rank and decrease the condition number of the MIMO channel. 
Due to the page limit, 
we will relegate the full analysis to the journal version of our work. 
Here,
we focus on a 2x2 MIMO channel as a toy example and plug \eqref{eq:Hfunc} into \eqref{eq:MIMOcapBase}, which yields \eqref{eq:HmatrixF} shown on the top of next page.
In \eqref{eq:HmatrixF},
%the equations  and \eqref{eq:Hmatrix} are combined to equation  to finally use in equation  as .
$\eta$ is a normalization factor of $\mathbf{H}$ and $\beta_{n_{ij}}$ is the random phase of re-radiated signal.

\begin{figure}[]
	\centering
	\includegraphics[width=1\linewidth]{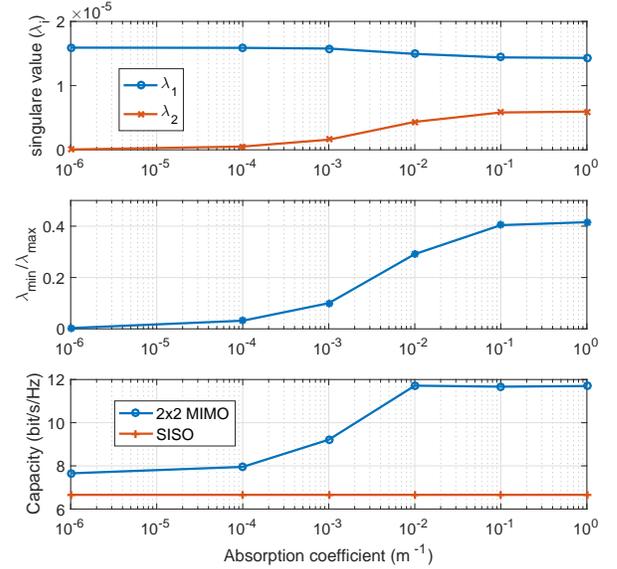}
		\caption{Singular value and condition number is affected by molecular absorption. Higher absorption leading to two times more capacity in compare with SISO}	
	\label{fig:analys}
\vspace{-0.3cm}
\end{figure}

To show the impact of the absorption coefficient on \eqref{eq:HmatrixF} and the MIMO capacity,
the channel transfer function and singular value are calculated for a practical range of absorption coefficient in a normal air condition at 60\,GHz,
while the channel distance is 50\,m and arrays are in parallel formation.
It should be noted that the actual value of absorption coefficient at 60\,GHz is around $2.7\times10^{-2}$.
In Figure~\ref{fig:analys}, we plot the results in terms of the singular value and the condition number. 
As one can observe,
for a very small absorption coefficient the largest singular value is much larger than the minimum one,
leading to an ill-conditioned MIMO channel matrix.
%This difference and the ratio can also be found in Figure~\ref{fig:analys}.
However,
as the absorption coefficient increases,
the singular values are getting closer and the inverse of condition number increases from zero toward one,
which implies a higher multiplexing gain.
Also, 
it can be seen from Figure~\ref{fig:analys} that the 2x2 MIMO capacity doubles that of a single-input-single-output (SISO) channel for a very large absorption coefficient.
%which achieves the maximum theoretical capacity because the minimum number of the transmitter and receiver antennas is 2.
In the next section,
more results for the mmWave spectrum with realistic absorption coefficients will be presented.
%%%%%%%%%%%%%%%%%%%%%%%%%%%%%%%%%%%%%%%%%%%%%%%%%%%%%%%%%%%%%%%%%%%%%%%%%
%%%%%%%%%%%%%%%%%%%%%%%%%%%%%%%%%%%%%%%%%%%%%%%%%%%%%%%%%%%%%%%%%%%%%%%%%

\section{Simulation and discussion}
\label{sec:simulSetup}

\subsection{Simulation set-up}
\label{sec:geometry}

\begin{figure}[h]
\centering
\includegraphics[width=0.7\linewidth]{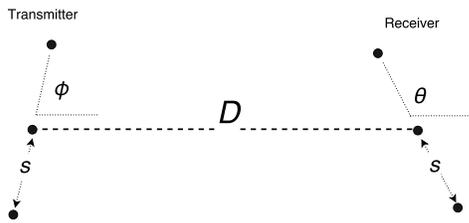}
\caption{A 3x3 MIMO system with uniform linear arrays}
\label{fig:geometry}
\end{figure}

To evaluate the MIMO capacity in mmWave and show the performance impact of the molecular absorption,
we consider a simple $n\times n$ MIMO system with Uniform Linear Arrays (ULAs),
where the inter-element spacing $s$ is equal to half of the wavelength at both transmitter and receiver,
and the channel distance is $D$.
The considered MIMO system is illustrated in Figure~\ref{fig:geometry}.
Moreover,
we consider uniform power allocation to transmitter arrays operating in an open-space LoS scenario.
The default values of the parameters are listed in Table~\ref{tab:simParam},
and different values will be explained when necessary.

In order to investigate the MIMO performance,
we evaluate the theoretical MIMO capacity with several configurations.
In the first step,
the parameters in Table~\ref{tab:simParam} are used.
More specifically,
the channel transfer matrix in \eqref{eq:Hmatrix} is obtained from \eqref{eq:Hfunc},
while the distance between each pair of transmitter-receiver antenna element is calculated by channel geometry.
Finally,
we compute the MIMO capacity using \eqref{eq:MIMOcapEig}.
%where both equations give the same result.
Since we apply random phases on NLoS components created by molecular re-radiation,
we conduct the evaluation of the MIMO capacity with molecular re-radiation for 5000 times and show the average result.

\begin{table}[h]
\centering
\caption{Simulation parameters}
\label{tab:simParam}
\begin{tabular}{|l|l|}
\hline
Transmitter and receiver distance ($D$) & $50$ m                       \\
Inter-element spacing ($s$)             & $0.5 \lambda$ (wave length)        \\
Transmitter arrays angle ($\phi$)          & $90^\circ $ \\
Receiver arrays angle ($\theta$)                   & $90^ \circ$ \\
Number of arrays on each side ($n$)     & $3$ \\
SNR & $20$ dB \\ \hline
\end{tabular}
\end{table}
%\subsection{Absorption Coefficient}

\begin{figure}
\centering
\includegraphics[width=\linewidth]{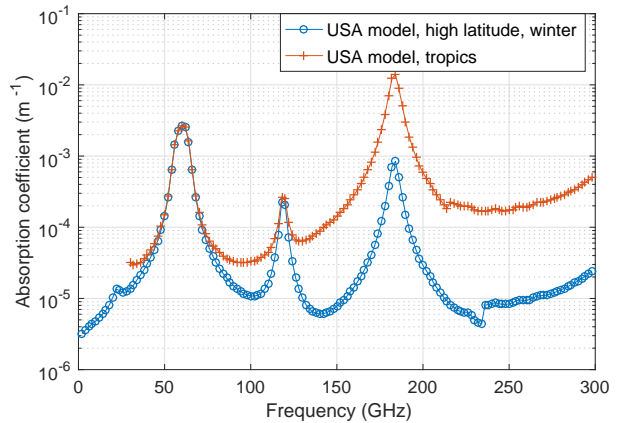}
\caption{The absorption coefficient in two different atmosphere. The temperature is $273$K and the pressure $1$atm.  }
\label{fig:kflog}
\vspace{-0.2cm}
\end{figure}

We use the online browsing and plotting tools\footnote{http://hitran.iao.ru/gasmixture/simlaunch},
which is based on HITRAN databases \cite{Rothman2012Database} to generate absorption coefficients for different single gas or some predefined standard gas mixture of atmosphere at sea level,
as shown in Table \ref{tab:gasMix}.
Since the Oxygen and water molecules play main roles in a normal air environment at mmWave bands and the Oxygen ratio is invariant,
we use the highest and lowest water ratio in Table~\ref{tab:gasMix},
i.e., the \emph{"USA model, high latitude, winter"} and \emph{"USA model, tropics"}.
The corresponding absorption coefficients in mmWave bands have been shown in Figure~\ref{fig:kflog} for an ambient temperature of $273$K and a sea level pressure of $1$atm.
It can be seen that there are three major absorption spikes in the mmWave spectrum as follows,
\begin{itemize}
  \item A pair of them appear at around 60\,GHz and 120\,GHz,
  which are attributed to Oxygen molecules.
  Note that the absorption coefficients are the same for both atmosphere cases,
  i.e., \emph{"USA model, high latitude, winter"} and \emph{"USA model, tropics"}.
  This is because the percentage of Oxygen is comparable for those two cases.
  \item The third one at 180\,GHz is created by water (\ce{H_2O}) molecules in the air.
  For a tropic atmosphere,
  the water ratio is higher than that of the winter atmosphere,
  and thus we can see a significant increase in terms of the absorption coefficient among these two atmosphere cases.
\end{itemize}
For comparison,
we also investigate a hypothetical vacuum case without absorption,
i.e., we set the the absorption coefficient to zero.

\begin{table*}[bt]
\centering
\caption{Atmosphere standard gas mixture ratio in percentage for different climates \cite{Rothman2012Database}}
\scriptsize
\label{tab:gasMix}
\begin{tabular}{|l|l|}
\hline
USA model, mean latitude, summer, H=0 & H2O: 1.860000   CO2: 0.033000   O3: 0.000003   N2O: 0.000032   CO: 0.000015   CH4: 0.000170   O2: 20.900001   N2: 77.206000 \\ \hline
USA model, mean latitude, winter, H=0 & H2O: 0.432000   CO2: 0.033000   O3: 0.000003   N2O: 0.000032   CO: 0.000015   CH4: 0.000170   O2: 20.900001   N2: 78.634779 \\ \hline
USA model, high latitude, summer, H=0 & H2O: 1.190000   CO2: 0.033000   O3: 0.000002   N2O: 0.000031   CO: 0.000015   CH4: 0.000170   O2: 20.900001   N2: 77.876781 \\ \hline
USA model, high latitude, winter, H=0 & H2O: 0.141000   CO2: 0.033000   O3: 0.000002   N2O: 0.000032   CO: 0.000015   CH4: 0.000170   O2: 20.900001   N2: 78.925780 \\ \hline
USA model, tropics, H=0               & H2O: 2.590000   CO2: 0.033000   O3: 0.000003   N2O: 0.000032   CO: 0.000015   CH4: 0.000170   O2: 20.900001   N2: 76.476779 \\ \hline
\end{tabular}
\end{table*}

\begin{figure*}
	\begin{center}
		\subfloat[]{\label{fig:Captropic}
\includegraphics[width=0.99\columnwidth ,clip=true, trim=0 0 0 0]{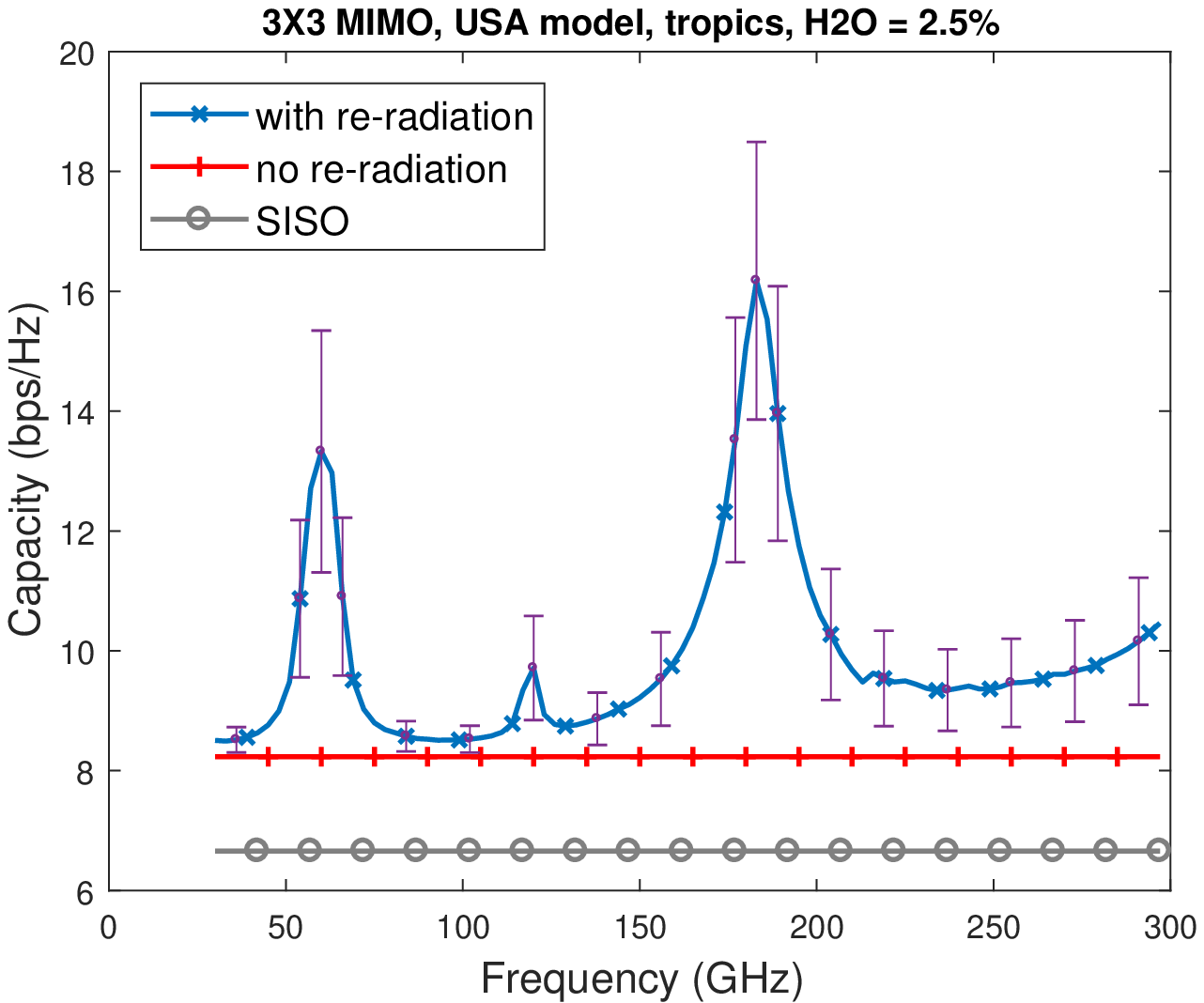}}
		\subfloat[]{\label{fig:Capwinter}
\includegraphics[width=0.99\columnwidth ,clip=true, trim=0 0 0 0]{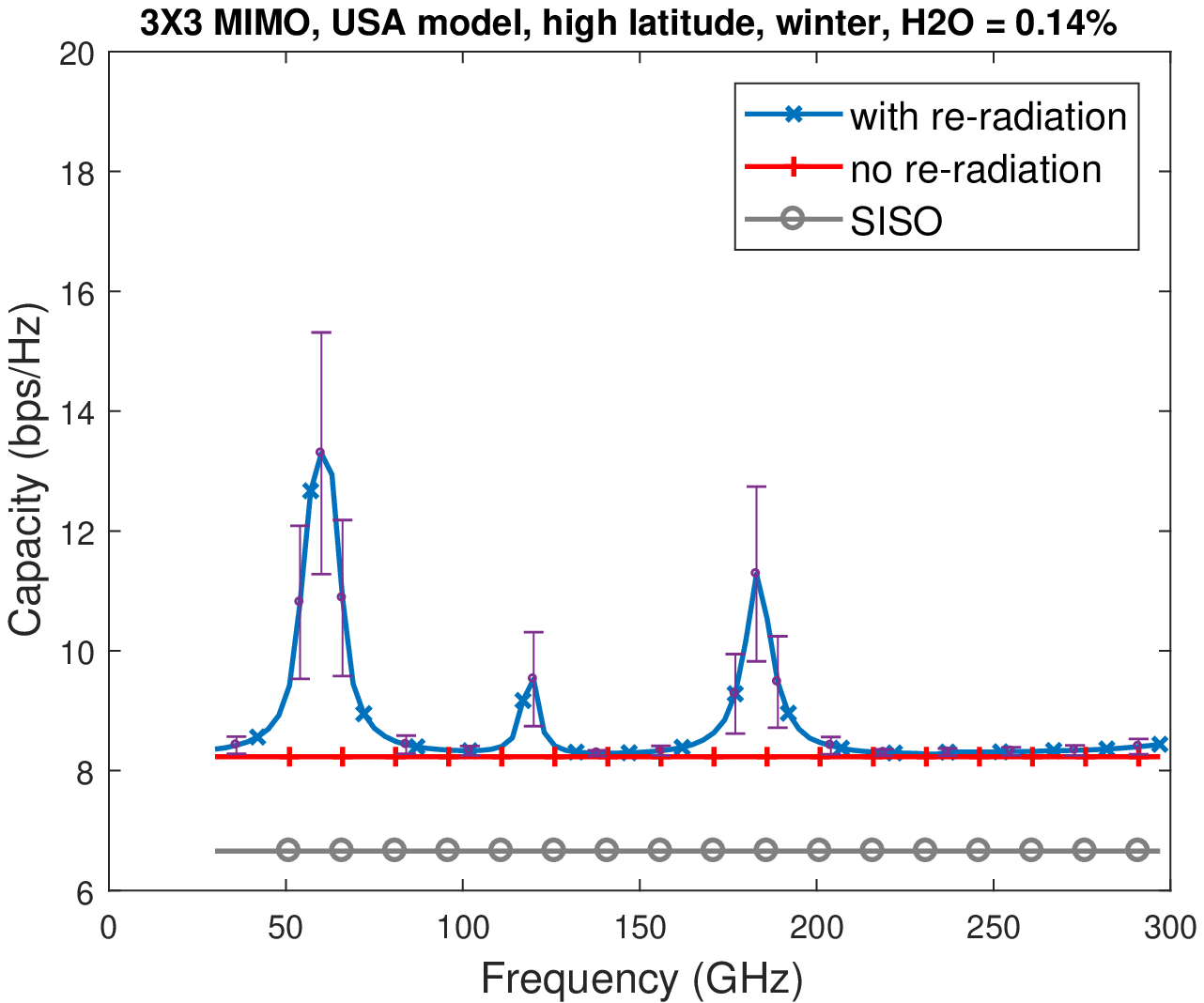}}
		\caption{3x3 MIMO capacity in the presence of molecular re-radiation (constant reception SNR)}	
		\label{fig:cap1}
	\end{center}
\vspace{-0.5cm}
\end{figure*}

\subsection{The MIMO Capacity vs. the Absorption Coefficient}\label{sec:simulResult}

In this subsection,
we first assume a constant reception SNR over the entire frequency spectrum,
and display the MIMO capacity in bit/sec/Hz for mmWave bands for a 3x3 MIMO with parallel ULA ($\phi=90^\circ$ and $\theta=90^\circ$) in Figure~\ref{fig:cap1}.
In this figure,
the MIMO capacity with absorption is compared with that without absorption.
As one can observe,
the capacity of the latter case is flat over the entire band and is not frequency selective,
which corroborates the common understanding up to now.
However,
when the absorption kicks in,
a significant increase shows up in certain frequency bands with high absorption.
For example,
the capacity is boosted by around 70\,\% at 60\,GHz for both atmosphere cases.
Furthermore,
since the tropic atmosphere contains more water molecules,
it leads to a considerable capacity increase at 180\,GHz in tropic atmosphere in comparison with those of the winter atmosphere and the hypothetical vacuum case without absorption. The confidence interval of 90\% also has been shown on capacity graphs in Figure~\ref{fig:cap1} in order to show the results distribution.

Numerically speaking,
we can calculate the capacity for a SISO channel,
which turns out to be 6.66 bit/sec/Hz.
According to the existing MIMO theory,
for a full-rank 3x3 MIMO channel with enough spatial diversity,
the theoretical capacity will be increased to $3 \times 6.66 \simeq 20$ bit/sec/Hz.
As discussed before,
the LoS MIMO channel suffers from poor spatial diversity and can achieve the maximum capacity only with some specific geometry configuration~\cite{sarris2005maximum},
which is not feasible for mobile communications.
Thus,
it can be seen the MIMO capacity without absorption is close to that of a SISO channel.
However,
if the molecular re-radiation is taken into account,
it can equivalently create a rich scattering environment,
and in turn increase the spatial diversity and the MIMO capacity,
as shown in~Figure \ref{fig:cap1}.

\begin{figure}
\centering
\includegraphics[width=\linewidth]{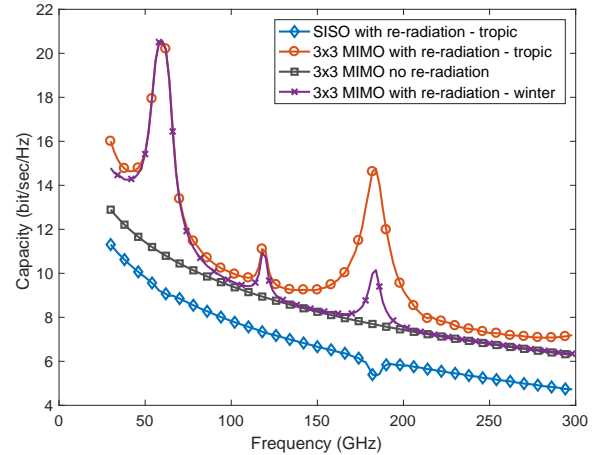}
\caption{3x3 MIMO capacity (constant transmitter power).}
\label{fig:cap2constantPow}
\vspace{-0.1cm}
\end{figure}

Note that in Figure \ref{fig:cap1},
we assume a constant reception SNR.
However,
the actual attenuation varies with the frequency because
(i) the free-space path loss increases with frequency,
and (ii) the molecular absorption also attenuates the signal.
Hence,
it is interesting to investigate whether the absorption attenuation will mitigate the MIMO performance improvement in high absorption frequency bands.
For this purpose,
a constant transmit power of 1\,W and a constant noise power of -100\,dBm are chosen in our next simulation.
Note that such parameter setting provides an SNR value similar to that was used in the previous step.
The results are exhibited in Figure~\ref{fig:cap2constantPow},
where the other parameter values are the same as those in Figure~\ref{fig:cap1}.
As one can observe,
the absorption attenuation has a marginal impact on the MIMO capacity improvement discussed before.
To show this,
we also simulate a SISO channel and plot its capacity in Figure~\ref{fig:cap2constantPow},
where we can see that the capacity slightly degrades at high absorption frequency bands at around 60\,GHz and 180\,GHz.
In summary,
the mmWave MIMO system can take advantage of the molecular absorption and re-radiation to generate more capacity,
which prevails the absorption attenuation.

\subsection{The MIMO capacity vs. The Antenna Number}
\label{sec:antenna-number}

\begin{figure}
\centering
\includegraphics[width=\linewidth]{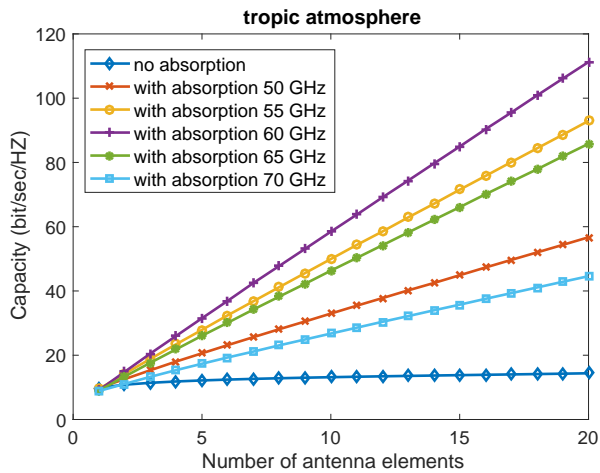}
\caption{The MIMO capacity increases linearly as the antenna number increases. 
}
\label{fig:cap3num}
\vspace{-0.4cm}
\end{figure}

As discussed before, 
one of the key benefits of the mmWave communication is its potential incorporation with the massive MIMO technology to generate a tremendous MIMO capacity. 
Hence, 
in this subsection, 
we investigate the molecular absorption effect on the MIMO capacity for a large number of antennas. 
Figure~\ref{fig:cap3num} displays the MIMO capacity as the number of antennas in both receiver and transmitter increases. 
All the simulation parameter values are based on Table~\ref{tab:simParam} except the arrays angles, 
i.e., $\theta$ and $\phi$, 
which are chosen randomly. 
Our simulation is repeated 5000 times and the average results have been shown for $\{50, 55, 60, 65,70\}$\,GHz. 

As can be seen from Figure~\ref{fig:cap3num}, 
with molecular absorption, 
the MIMO capacity increases linearly as the number of antennas increases. 
Such performance gain is much more obvious for $55\sim65$\,GHz around the natural resonance frequency of Oxygen, 
which creates an opportunistic spectrum window for high-efficiency MIMO communications in mmWave.
Also note that without molecular absorption, 
the MIMO capacity is frequency non-selective and does not increase with the number of antennas due to the dominant LoS transmissions.
Such contrast comparison can be easily seen from Figure~\ref{fig:cap3num}.

\balance
%\vspace{0.1cm}
\section{Conclusion}
\label{sec:con}
In this paper,
we investigated the MIMO capacity for mmWave,
where the gas molecules in our atmosphere like Oxygen and water not only can increase the path loss,
but also can re-radiate a copy of signal wave and equivalently create multi-path channels.
Our results showed that in certain frequency bands,
where the absorption coefficient is significantly high,
the MIMO channel can achieve a capacity that is close to the theoretical limit,
thanks to the great improvement in spatial diversity.
Our new discovery fundamentally changes our understanding on the relationship between the MIMO capacity and the frequency spectrum,
especially for mmWave communications with massive MIMO.
%Although, we exploit well-known molecular absorption model in our simulation but for the future work, real measurement should be taken into account to prove the simulation results.
%\vspace{0.1cm}
\bibliographystyle{IEEEtran}
\bibliography{Ref}
\end{document}